\documentclass[superscriptaddress,twocolumn]{revtex4}
\usepackage{amssymb}
\usepackage[tbtags]{amsmath}
\usepackage{graphicx}
\usepackage{epsfig,graphicx,times}
\usepackage{amsbsy}
\usepackage{latexsym,epsfig,graphicx}
\usepackage{dcolumn}
\usepackage{subfigure}
\usepackage{comment}
\usepackage{color}
\usepackage[colorlinks,urlcolor=blue,citecolor=blue]{hyperref}
\usepackage{amstext}
\usepackage{amssymb}
\usepackage{setspace}
\usepackage{lipsum}
\usepackage{mathtools}
\usepackage{floatrow}
\usepackage[T1]{fontenc}

\begin{document}
\title{Unidirectional transmission of single photons under non-ideal chiral photon-atom interactions}
\author{Cong-Hua Yan\footnote{yanconghua@126.com}}
\affiliation{College of Physics and Electronic
Engineering, Sichuan Normal University, Chengdu 610068, China}
\author{Ming Li}
\affiliation{Key Laboratory of Quantum Information, University of Science and
Technology of China, CAS, Hefei, Anhui 230026, China}
\author{Xin-Biao Xu}
\affiliation{Key Laboratory of Quantum Information, University of Science and
	Technology of China, CAS, Hefei, Anhui 230026, China}
\author{Yan-Lei Zhang}
\affiliation{Key Laboratory of Quantum Information, University of Science and
Technology of China, CAS, Hefei, Anhui 230026, China}
\author{Hao Yuan}
\affiliation{Key Laboratory of Opto-Electronic Information Acquisition and Manipulation of Ministry of Education, School of Physics and Material Science, Anhui University, Hefei 230601, China}
\author{Chang-Ling Zou\footnote{clzou321@ustc.edu.cn}}
\affiliation{Key Laboratory of Quantum Information, University of Science and
Technology of China, CAS, Hefei, Anhui 230026, China}
\affiliation{State Key Laboratory of Quantum Optics and Quantum Optics Devices, Shanxi University, Taiyuan 030006, China}
\date{\today}

\begin{abstract}
Single-photon transport in non-ideal chiral photon-atom interaction structures generally contains information backflow and thus limits the capabilities to transfer information between distant emitters in cascaded quantum networks. Here, in the non-ideal chiral case, a $V$-type atom coupled to a waveguide is proposed to realize completely unidirectional transmission of the single photons in a superposition state of different frequencies. A microwave field is introduced to drive the two excited states of the atom and results in photon conversion between two transitions. By adjusting the Rabi frequency and the phase of the external driving field, the transport behaviors of incident photons with specific frequencies can be optimized to complete transmission or reflection. Based on the constructive interferences between photons from different traveling paths, the transmission probabilities of the specific-frequency photons could be enhanced. Due to photon conversions with compensating the dissipations and ensuring the complete destructive interference, the ideal unidirectional transmission contrast can be maintained to $\pm 1$, even when the atom has dissipations into the non-waveguide modes.
\end{abstract}
\maketitle

\section{Introduction}
Controlling the direction of photon scattering is central to modern photonics applications ranging from quantum networks to quantum-information processing~\cite{KimbleN1023,LodahlRMP347,ChangNP685}. Typically, a cascaded quantum network~\cite{GardinerPRL2269,CarmichaelPRL2273} consisting of multiple emitters coupled to a one-dimensional reservoir requires the unique feature that excitations can only propagate along a single direction, and then drives successive systems in the network in a unidirectional way~\cite{StannigelNJP063014}. Therefore, nanophotonic waveguides utilized to be the reservoirs are well suited for these applications as they confine photons to a one-dimensional geometry with only forward and backward directions and thereby increase the photon-emitter interaction with constituting a deterministic photon-emitter interface~\cite{GuPR1,RoyRMP021001}. However, quantum emitters interacting equally with photons in either of the two propagation directions along a regular waveguide can be served as quantum mirrors~\cite{ShenPRL213001} and quantum beats~\cite{ZhengPRL113601} and inevitably contain information backflow, which suppress the fidelity of quantum state transfer between different emitters in cascaded quantum networks~\cite{LodahlN473}.

\begin{figure}[htbp]
  \centering
  \includegraphics[width=7.0cm]{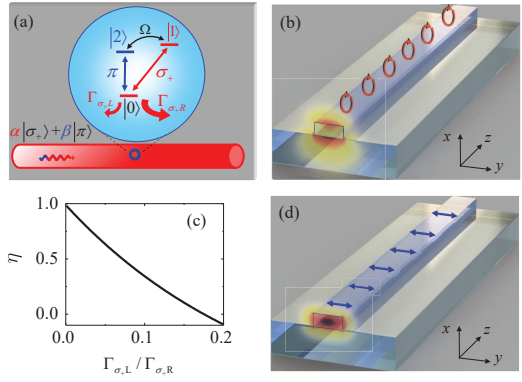}\\
  \caption {A $V$-type atom with $\sigma_+$ and $\pi$ transitions coupled to a waveguide under non-ideal chiral interactions. (a) The waveguide sustains two different modes of $\sigma_{+}$ and $\pi$. A single photon in the superposition state of different frequencies $\omega_{\sigma_{+}}$ and $\omega_\pi$ as $\alpha|{\sigma_{+}}\rangle+\beta|\pi\rangle$ with $|\alpha|^2+|\beta|^2=1$ and $\beta=e^{i\psi}\sqrt{1-|\alpha|^2}$ depicted as a wiggly wave is injected from the left side of the waveguide and excites the corresponding waveguide modes. An external microwave field with the Rabi frequency of $\Omega$ is coupled to the exited states $|1\rangle$ and $|2\rangle$. Although the $\sigma_{-}$ transition is also involved in the atom, we can utilize the AC Stark effect
to adjust the corresponding energy level of the atom by applying a strong far-detuned laser (with the $\sigma_-$ circularly polarization) to realize a spin-selective energy shift~\cite{YangNC12613,WilkinsonAPL133104}. Consequently, the microwave field only couples to levels$|1\rangle$ and $|2\rangle$. (b) The photon with the frequency of $\omega_{\sigma_{+}}$ (simplified as ${\sigma_{+}}$ photon) resonantly exciting the transverse-magnetic (TM) mode of the rectangular waveguide interacts with the ${\sigma_{+}}$ transition of $|0\rangle\leftrightarrow|1\rangle$ under non-chiral interactions of different decaying rates as $\Gamma_{{\sigma_{+}} R}>\Gamma_{{\sigma_{+}} L}$. (c) The unidirectional transmission contrast (UTC) (i.e., $\eta$) of the ${\sigma_{+}}$ photon is suppressed by decreasing the chirality of $\Gamma_{{\sigma_{+}} L}/\Gamma_{{\sigma_{+}} R}$. (d)  The photon with the frequency of $\omega_\pi$ (simplified as $\pi$ photon) resonantly excites the transverse-electric (TE) mode of the rectangular waveguide and equally couples to the $\pi$ transition of $|0\rangle\leftrightarrow|2\rangle$ with $\Gamma_{\pi R}=\Gamma_{\pi L}$. Here, the $z$ axis coincides with the waveguide axis and indicates the propagation direction of the photon along the waveguide. The $y$ axis is chosen to be the quantization axis, wherein the waveguide is different from free space optics and the electric field gradient along $y$ axis leads to a $z$ component~\cite{LodahlN473}. Consequently, the circularly polarization mode $\sigma_{+}$ is in the $x-z$ plane and the $\pi$ polarized mode is parallel to the $y$ axis~\cite{PetersenS67}. }\label{fig1}
\end{figure}

Recently, chiral photon-emitter interaction originated from the optical spin-orbit coupling~\cite{BliokhNP796} constitutes an exciting new approach to controlling the transporting direction of single photons~\cite{LodahlN473,MitschNC5713,JungePRL213604,PetersenS67,ColesNC11183,SolnerNN775,
SayrinPRX041036,ScheucherS1577}.
 When a spin-momentum-locked light transversely confined in a nanophotonic waveguide is coupled to a quantum emitter with polarization-dependent dipole transitions, then direction-dependent emission, scattering, and absorption of photons are obtained. It indicates that a right-hand circularly polarized dipole emitter is matched to the local polarization of the waveguide and then emits solely along the  right direction with the transmission probabilities of $T=1$ and reflection probabilities $R=0$. Therefore, the unidirectional transmission contrast (UTC) defined as  $\eta=(T-R)/(T+R)$~\cite{ShenPRL173902,XiaPRA043802} in the ideal chiral case for a photon injected from the left to right side of the waveguide is $\eta=1$.
Based on the unidirectional transporting characters, photons emitted by the first emitter will be completely absorbed by the subjacent emitter without information backflow. It means that the ideal chiral photon-emitter interactions with perfectly unidirectional transmission allow deterministic state transfer between emitters along the waveguide. Such an approach paves the way to engineering the large scale cascaded quantum networks~\cite{PichlerPRA042116,MahmoodianPRL240501,BorregaardAQT1800091,GuimondPRA033829,GrankinPRA043825}, which is in great demand in quantum information processing.

Practically, the chiral photon-atom interactions in realistic systems are complicated, being highly sensitive to the effects of imperfect directional couplings~\cite{HurstNL5475}, the non-perfectly circular polarizations of the photonic fields~\cite{TangPRA043833}, and the positions of the emitters~\cite{DowningPRL057401}. For example, though a quantum dot (QD) placed at a singular position of the glide-plane photonic crystal waveguide known as the C-point can display a spin-dependent unidirectional emission, the corresponding photon-QD interaction is relatively small~\cite{LangJO045001}. This forces a compromise between strong photon-emitter interactions and makes those interactions with high chiralities. Experimentally, the decaying rates of the excited atom to the right and left directions (i.e.,  $\Gamma_R$ and $\Gamma_L$) can be optimized to $\Gamma_R/\Gamma_L\approx5$ with most of radiation photons decaying into the glide-plane photonic crystal waveguide~\cite{MahmoodianOME43}. Additionally, perfect spin-orbit couplings require emitters exactly resonating with the circular polarization of surface plasmon-polariton modes (SPPMs) in propagating along the nanowire, and hence the practical atom with the transition frequency far from the SPPMs inevitably reflects SPPMs~\cite{KornovanPRB115162}. It means that the UTC in non-ideal chiral systems is always less than unity as $\eta<1$.
Consequently, the realistic systems inevitably including reverse processes of photon transport lead to the dipole-dipole coupling between emitters with repeated exchange of excitations, which suppress the abilities of photons as flying qubits and two-level emitters as stationary qubits for distributing quantum information carried by the circularly polarized photon in a quantum network~\cite{LodahlN473}. Therefore, how to realize completely unidirectional transmission of single photons in non-ideal chiral photon-atom interaction structures is still a challenge.

In this work, we concentrate on the improvement of the UTC in the non-ideal chiral photon-atom interactions with a $V$-type atom coupled to the waveguide. A full quantum theory in real space is adopted to investigate the transport properties of a single photon with a superposition state of different frequencies. By properly modulating the Rabi frequency and the phase of the external field driven on the atom, the UTC of the incident photon with a specific frequency can be optimized to complete transmission or reflection, even when the atom has dissipations into the non-waveguide modes.

This paper is organized as follows. Our model is presented in Sec. II, wherein an
effective real-space Hamiltonian of the system is introduced and the exact transmission spectra
of one-photon solution are given. In Sec. III we numerically analyze how the external driving field
can be utilized to modulate the photon transport. Finally, in Sec. IV, we conclude our work and suggest
experimental demonstrations of our proposal with the current photonic techniques.

\section{Model and solutions}

As shown in Fig.~1(a), we consider an atom that consists of $V$-type levels coupled to a waveguide, with a ground state $|0\rangle$ initially prepared in spin-up~\cite{AtatureS551,PressN218,CarterNP329} and two excited states $|1\rangle$ and $|2\rangle$ at frequencies $\omega_g$, $\omega_{e_{\sigma_{+}}}$, and $\omega_{e_\pi}$. In traditional studies, only one waveguide mode is used to couple with the atom. In this work, one rectangular waveguide with two different modes  simultaneously interacting with the atom is proposed to realize completely unidirectional transmission of the single photons. Figure~1(b) illustrates the transverse-magnetic (TM) mode of the rectangular waveguide, which exhibits strong charity due to the gradient of the electric field on top of the waveguide~\cite{JungePRL213604}. It couples with the $\sigma_{+}$ transition of the atom and can be used for near unidirectional transport of single photons. However, the chirality of the TM mode is non-ideal and is also position-dependent. As shown by Fig.~1(c), the UTC decreases with nonideality of the chirality. To overcome this problem, as illustrated in Fig.~1(d), we use a transverse-electric (TE) mode of the rectangular waveguide to couple with the $\pi$ transition of the atom. By preparing the incident single photon in the superposition state of two photonic modes TE and TM as $\alpha|{\sigma_{+}}\rangle+\beta|{\pi}\rangle$ with $|\alpha|^2+|\beta|^2=1$ and $\beta=e^{i\psi}\sqrt{1-|\alpha|^2}$~\cite{KoshinoPRA010301}, and convert the corresponding frequencies between $\omega_{\sigma_{+}}$ and $\omega_\pi$ to match the atomic transitions,  the single-photon transport can be controlled by a microwave field with Rabi frequency $\Omega$ and phase $\phi$ that couples the two internal excited states of the atom. Typically, the TE mode that interacts with the ${\sigma_{+}}$ transition of the atom has a high degree of chiral photon-atom interaction with $\Gamma_{{\sigma_{+}} R}>\Gamma_{{\sigma_{+}} L}$, while the TM mode propagating in two directions couples equally to the $\pi$ transition with $\Gamma_{\pi R}=\Gamma_{\pi L}$, respectively.

The system can be described by the following Hamiltonian in the real space under the rotating wave approximation with $\hbar=1$~\cite{ShenPRL213001,ShenPRA023837}:
\begin{eqnarray}
\label{eq:1}
H&=&\int dz\sum_{j={\sigma_{+}},\pi}\bigg\{c^\dag_{j,R}(z)\bigg(-iv_g\frac{\partial}{\partial z}-\omega_{e_{j}g}+i\gamma_j\bigg)\nonumber\\
&&c_{j,R}(z)
+c^\dag_{j,L}(z)\bigg(iv_g\frac{\partial}{\partial z}-\omega_{e_{j}g}+i\gamma_j\bigg)c_{j,L}(z)\nonumber\\
&+&V_{j R}\delta(z)[c^\dag_{j,R}(z)a_{g}^\dag a_{e_j}+c_{j,R}(z)a_{e_j}^\dag a_{g}]\nonumber\\
&+&V_{j L}\delta(z)[c^\dag_{j,L}(z)a_{g}^\dag a_{e_j}+c_{j,L}(z)a_{e_j}^\dag a_{g}]\bigg\}\nonumber\\
&+&\Omega[e^{i\phi} a_{e_{\sigma_{+}}}^\dag a_{e_\pi}+e^{-i\phi} a_{e_\pi}^\dag a_{e_{\sigma_{+}}}],
\end{eqnarray}
wherein, $c^\dag_{j,R}(z)$ ($c^\dag_{j,L}(z)$) is the creation operator for right-going (left-going) photon of the frequency $\omega_j$ at the $z$ position of the waveguide. For simplicity, we linearize the waveguide dispersion relation in the vicinity of the two transition frequencies. The group velocity $v_g$ for photons with different frequencies is the same. $\omega_{e_jg}=\omega_{e_j}-\omega_g$ is the atom transition frequency between levels $|1\rangle$, $|2\rangle$, and $|0\rangle$. $\gamma_j$ is the dissipation rates of the atom resulted from the atom coupling to the reservoir~\cite{ShenPRA023837}. $V_{jR(L)}$ is the photon-atom coupling strength of the photon propagating along the right (left) direction, with $\Gamma_{jR(L)}=V_{jR(L)}^2/v_g$ being the atomic decaying rate to the waveguide. As we are only interested in a narrow range in the vicinity of the atomic resonant frequency, $V_{jR(L)}$ is safely assumed to be independent of frequency~\cite{ZhengPRL113601,ZhouPRl100501,ZhouPRL103604}.

In this work, we assume that the atom is originally prepared at the ground state of $|0\rangle$, and the incident single photon of the superposition state is injected from the left of the waveguide. During the scattering process, the incident photon may either remain at the same frequency or convert to the other frequency. Therefore, the general eigenstate of the system should take the following form:

\begin{eqnarray}
\label{eq:wave}
|\Psi\rangle&=&\sum_{j={\sigma_{+}},\pi}\bigg\{\int dz[\phi_{kj,R}c^\dag_{j,R}(z)+\phi_{kj,L}c^\dag_{j,L}(z)]|0_{\text{w}},0\rangle\nonumber\\
&+&e_{aj}a_{ej}^\dag a_g|0_{\text{w}},g\rangle\bigg\},
\end{eqnarray}
where $|0_{\text{w}},0\rangle$ is the vacuum, with zero photon in the waveguide and the atom in the ground state $|0\rangle$. $e_{aj}$ is the excitation amplitude of the atom in the state $|1\rangle$ or $|2\rangle$. The spatial dependence of wavefunctions can be expressed as:

\begin{eqnarray}
\label{eq:1}
\phi_{k{\sigma_{+}},R}&=&{\alpha} e^{ik_{\sigma_{+}} z}\theta(-z)+t_{\sigma_{+}} e^{ik_{\sigma_{+}} z}\theta(z),
\end{eqnarray}
\begin{eqnarray}
\label{eq:1}
\phi_{k{\sigma_{+}},L}&=&r_{\sigma_{+}} e^{-ik_{\sigma_{+}} z}\theta(-z),
\end{eqnarray}
\begin{eqnarray}
\label{eq:1}
\phi_{k\pi,R}&=&\beta e^{ik_\pi z}\theta(-z)+t_\pi e^{ik_\pi z}\theta(z),
\end{eqnarray}
\begin{eqnarray}
\label{eq:1}
\phi_{k\pi,L}&=&r_\pi e^{-ik_\pi z}\theta(-z),
\end{eqnarray}
wherein $\theta(z)$ is the step function. $k_j=\omega_j/v_g$, with $\omega_j$ being the frequency of the different photons along the waveguide. $T_j=|t_j|^2$ and  $R_j=|r_j|^2$ give the transmission and reflection probabilities of different frequencies, respectively. Following the procedure in Refs.~\cite{ShenPRL213001,ShenPRA023837,YanPRA045801,YuanOL5140} that directly deals with the photon scattering eigenstates in real space, we can solve the time-independ Schrodinger equation $H|\Psi\rangle=\Delta|\Psi\rangle$ with $\Delta=\omega_j-\omega_{e_jg}+i\gamma_j$, and obtain the following formulas:
\begin{eqnarray}
\label{eq:rsigma}
r_{\sigma_{+}}&=&\frac{\alpha\sqrt{\Gamma_{{\sigma_{+}} L}\Gamma_{{\sigma_{+}} R}}(i\Delta-\Gamma_\pi)}
{(i\Delta-\Gamma_{{\sigma_{+}}})(i\Delta-\Gamma_{\pi})
+\Omega^2}\nonumber\\
&+&\frac{i\beta\Omega e^{i\phi}\sqrt{\Gamma_{\pi R}\Gamma_{{\sigma_{+}} L}}}{(i\Delta-\Gamma_{{\sigma_{+}}})(i\Delta-\Gamma_{\pi})
+\Omega^2},
\end{eqnarray}
\begin{eqnarray}
\label{eq:tsigma}
t_{\sigma_{+}}&=&\alpha+\sqrt{\frac{\Gamma_{{\sigma_{+}} R}}{\Gamma_{{\sigma_{+}} L}}}r_{\sigma_{+}}\nonumber\\
&=&\alpha+\frac{\alpha\Gamma_{{\sigma_{+}} R}(i\Delta-\Gamma_{\pi})}{(i\Delta-\Gamma_{{\sigma_{+}}})(i\Delta-\Gamma_{\pi})
+\Omega^2}\nonumber\\
&+&\frac{i\beta\Omega e^{i\phi}\sqrt{\Gamma_{{\sigma_{+}} R}\Gamma_{\pi R}}}{(i\Delta-\Gamma_{{\sigma_{+}}})(i\Delta-\Gamma_{\pi})
+\Omega^2},
\end{eqnarray}
\begin{eqnarray}
\label{eq:rpi}
r_\pi&=&\frac{\beta\sqrt{\Gamma_{\pi L}\Gamma_{\pi R}}}{i\Delta-\Gamma_{\pi}}+\frac{i\Omega e^{-i\phi}\sqrt{\Gamma_{\pi L}/\Gamma_{{\sigma_{+}} L}}}{i\Delta-\Gamma_{\pi}}r_{\sigma_{+}},
\end{eqnarray}
\begin{eqnarray}
\label{eq:tpi}
t_\pi&=&\beta+\sqrt{\frac{\Gamma_{\pi R}}{\Gamma_{\pi L}}}r_\pi\nonumber\\
&=&\beta+\frac{\beta\Gamma_{\pi R}}{i\Delta-\Gamma_{\pi}}+\frac{i\Omega e^{-i\phi}\sqrt{\Gamma_{\pi R}/\Gamma_{{\sigma_{+}} L}}}{i\Delta-\Gamma_{\pi}}r_{\sigma_{+}},
\end{eqnarray}
with $\Gamma_{\sigma_{+}}=\frac{\Gamma_{{\sigma_{+}} L}+\Gamma_{{\sigma_{+}} R}}{2}$ and $\Gamma_\pi=\frac{\Gamma_{\pi L}+\Gamma_{\pi R}}{2}$.

The results in the above formulas satisfy $\sum_{j={\sigma_{+}},\pi} (T_j+R_j)=1$ with $\gamma_j=0$, as required by probability conservation. Since the chiral photon-atom interaction is based on the ${\sigma_{+}}$ transition of the atom, we mainly concentrate on the transmission unidirectional contrast of the ${\sigma_{+}}$ photon as $\eta_{\sigma_{+}}=(T_{\sigma_{+}}-R_{\sigma_{+}})/(T_{\sigma_{+}}+R_{\sigma_{+}})$. Generally, the value of phase $\phi$ is related to the choice of the time origin and  can be set with arbitrary amplitudes~\cite{FangPRL153901}. Consequently, the phase $\phi$ by itself usually has no measurable physical results in transmission spectra of the single atom system~\cite{YuanOL5140}. In contrast, when we focus on the structure illustrated in Fig.~1 with a $V$-type three-level atom interacted with a single photon in the superposition state, the phase $\phi$ of the external driving field has direct effects on the transmission and reflection spectra, which can be seen obviously in the above formulas. Significantly, Eq.~(\ref{eq:rsigma}) indicates that the reflection amplitudes include two kinds of sources as photons decay to the left direction from either the ${\sigma_{+}}$ transition directly or from the conversion of the $\pi$ photons. Meanwhile, the transmission amplitudes in Eq.~(\ref{eq:tsigma}) include three sources as photons transmitting directly along the waveguide, photons decaying to the right direction from the ${\sigma_{+}}$ transition, and photons decaying from the conversion of the $\pi$ photons. Especially, Eq.~(\ref{eq:tsigma}) indicates that the difference of amplitudes between the two kinds of decaying photons can be modulated to either positive or negative, and results in constructive or destructive interferences respectively among the photons of different processes. Therefore, the transporting properties of ${\sigma_{+}}$ photons in the present structure depend on the two-source and three-source interferences and will be demonstrated in the rest of the work.

\section{Unidirectional transmission of single photons controlled by the external field}
\begin{figure}[htbp]
  \centering
  \includegraphics[width=8cm]{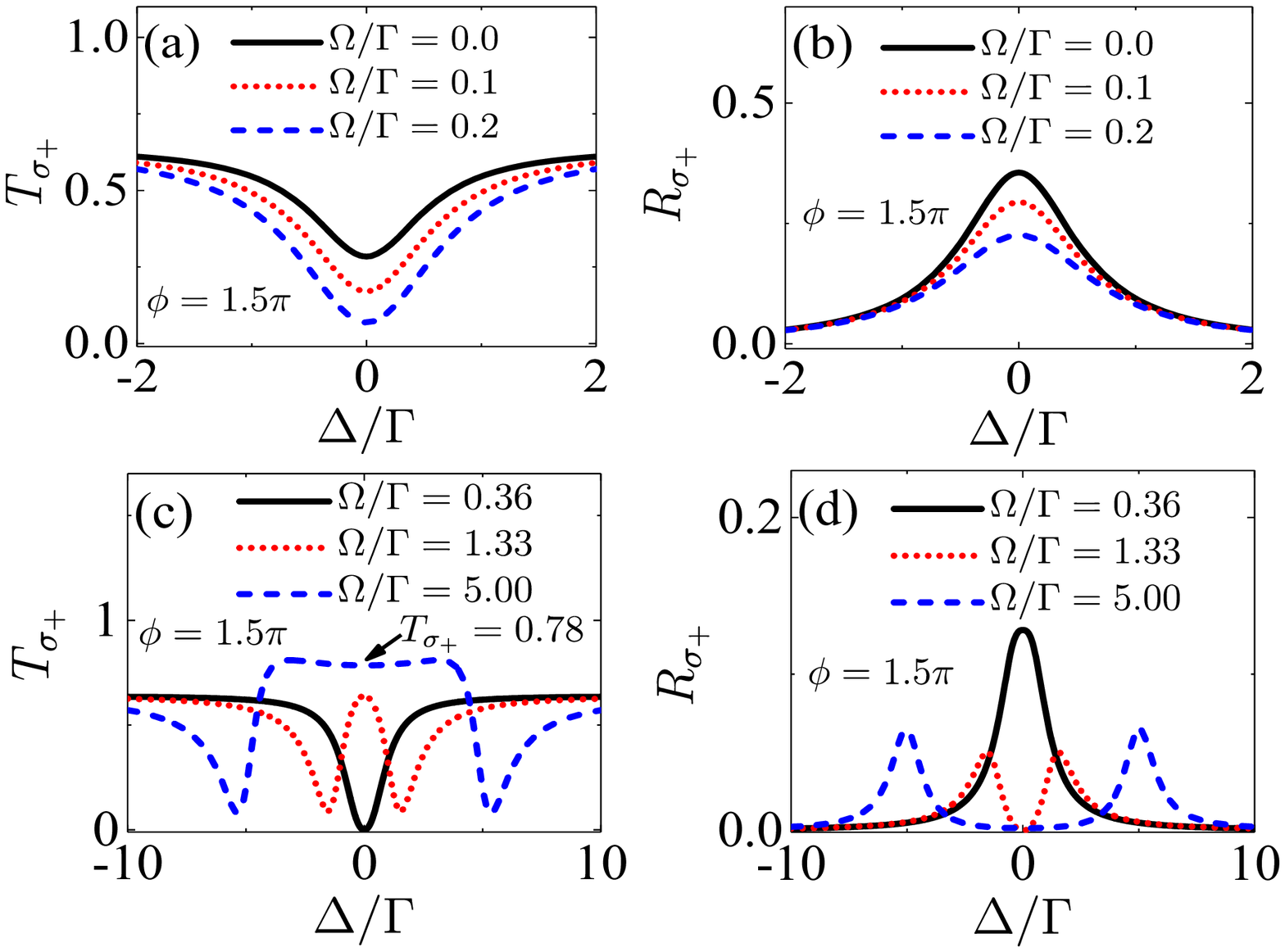}\\
  \caption {Transmission and reflection spectra of ${\sigma_{+}}$ photons controlled by the Rabi frequency of the external driving field, with $\alpha=0.8,\,\psi=0,\,\Gamma_{{\sigma_{+}} R}=\Gamma,\,\Gamma_{{\sigma_{+}} L}=0.2\Gamma,\,\phi=1.5\pi,$ and $\gamma_{\sigma_{+}}=\gamma_\pi=0$. $T_{\sigma_{+}}=0$ and $R_{\sigma_{+}}>0$  indicate that the ${\sigma_{+}}$ photon is completely reflected to the left direction of the waveguide when $\Omega=0.36\Gamma$, while the behaviors of $\Omega=1.33\Gamma$ are opposite. Therefore, the transmission directions of the ${\sigma_{+}}$ photon can be controlled by adjusting $\Omega$.}\label{fig1}
\end{figure}

Firstly, we investigate the transmission and reflection probabilities influenced by the Rabi frequency $\Omega$ of the driving field with $\alpha=0.8$, $\psi=0$, $\Gamma_{{\sigma_{+}} R}=\Gamma$, $\Gamma_{{\sigma_{+}} L}=0.2\Gamma$~\cite{MahmoodianOME43}, $\Gamma_{\pi R}=\Gamma_{\pi L}=\Gamma$, $\phi=1.5\pi$, and $\gamma_j=0$. As plotted in Fig.~2(a), in the absence of driving field ($\Omega=0$), the non-ideal chiral photon-atom interaction with $\Gamma_R/\Gamma_L=5$ results in photon transporting bidirectionally with $T_{\sigma_{+}}<R_{\sigma_{+}}$. It indicates that in this non-ideal chiral case, the information carried by the incident photon will be reflected to a great extent. By increasing the driving field with $\Omega>0$, as shown in Figs.~2(a) and 2(b), both the transmission and reflection probabilities are suppressed. These suppressings come from the fact that the driving field coupled to the two excited states not only changes the amplitudes of the photons decaying from the ${\sigma_{+}}$ transition but also introduces the other source of photons by converting the $\pi$ photons to the ${\sigma_{+}}$ photons. Consequently, the three-source and two-source interferences are reconstituted and then the corresponding transmission and reflection probabilities are modified respectively.

Especially, when $\Omega=0.36\Gamma$, as shown in Fig.~2(c), the transmission probability of the ${\sigma_{+}}$ photon is zero.
It can be directly calculated from Eq.~(\ref{eq:tsigma}) that when $\phi=1.5\pi$ and $\Omega=(\beta \sqrt{\Gamma_{{\sigma_{+}} R}\Gamma_{\pi R}}+\sqrt{\beta^2\Gamma_{{\sigma_{+}} R}\Gamma_{\pi R}
-4\alpha^2\Gamma_{\sigma_{+}}\Gamma_\pi+4\alpha^2\Gamma_{{\sigma_{+}} R}\Gamma_{\pi}})/2\alpha$, the total amplitudes of photons from two sources decaying to the right direction satisfy the relation of $\sqrt{\Gamma_{{\sigma_{+}} R}/\Gamma_{{\sigma_{+}} L}} r_{\sigma_{+}}=-\alpha$.
It indicates that the amplitudes of the reemitted photon are modulated to the same value as that of the direct-transmitted photon but differ with a $\pi$ phase and consequently construct a perfect destructive interference of photons from three different sources with $T_{\sigma_{+}}(\Delta=0,\,\Omega=0.36\Gamma)=0$~\cite{RosenblumPRA033814}. In the present of strong external field of $\Omega>\Gamma$, the transmission spectra are split into two dips. This splitting is the effect of Rabi splitting induced by the external field~\cite{YuanOL5140}.
When $\phi=1.5\pi$ and the driving field is strong enough as $\Omega=5\Gamma$, the decaying amplitudes of photons to the right direction from the conversion with positive values can be larger than those of photons decaying from the ${\sigma_{+}}$ transition and thus $t_{\sigma_{+}}>\alpha$. It results in a constructive interference between the directly transmitted photons and the atomic radiation photons with the transmission probabilities of the resonant ${\sigma_{+}}$ photon being enhanced with $T_{\sigma_{+}}(\Delta=0,\,\Omega=5.0\Gamma)=0.78>|\alpha|^2=0.64$, as shown in Fig.~2(c).

Similarly, Eq.~(\ref{eq:rsigma}) indicates that adjusting the driving field as $\Omega=\alpha\sqrt{\Gamma_{{\sigma_{+}} L}}\Gamma_{\pi}/i\beta e^{i\phi}\sqrt{\Gamma_{\pi R}}$ leads to the happening of complete destructive interference between two sources of the left-decaying photons with $R_{\sigma_{+}}(\Delta=0,\,\Omega=1.33\Gamma)=0$ in Fig.~2(d).
If the driving field is adjusted to much stronger than the atomic decaying rates of ${\sigma_{+}}$ transition to the left direction as $\Omega\gg\Gamma_{{\sigma_{+}}_L}$, then the reflection amplitude in Eq.~(\ref{eq:rsigma}) is suppressed to $r_{\sigma_{+}}=0$, which can be seen from the Fig.~2(d) of $R_{{\sigma_{+}}}(\Delta=0,\,\Omega=5.0\Gamma)=0$.
 Therefore, the transmission properties of the ${\sigma_{+}}$ photon can be controlled by adjusting the Rabi frequencies of the driving field.

\begin{figure}[htbp]
  \centering
  \includegraphics[width=8cm]{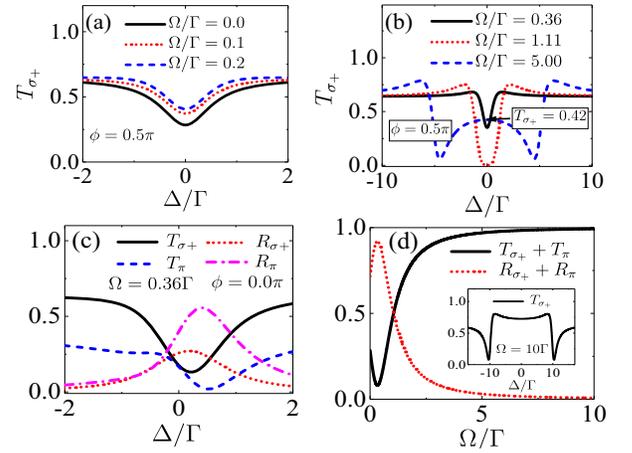}\\
  \caption {Transmission and reflection spectra of ${\sigma_{+}}$ and $\pi$ photons. (a) to (c) Transporting properties modulated by the phase of the external field. (d) Transporting properties controlled by the Rabi frequency of the driving field with $\Delta=0$, $\phi=1.5\pi$. Other parameters are the same as in Fig.~2.}\label{fig1}
\end{figure}

Next, we investigate the transmission properties affected by the phase of the driving field. As shown in Fig.~3(a), when $\phi=0.5\pi$, the transmission probabilities increase by enhancing the driving field, which are contrast to the behaviors of transmission spectra shown in Fig.~2(a). Practically, there is a complete destructive interference between the reemitted photons and the directed transmission photons with $T_{\sigma_{+}}=0$ under the Rabi frequency of $\Omega=1.11\Gamma$, as plotted in Fig.~3(b). We also find that the maximum transmission properties of the resonant photon ($T_{\sigma_{+}}=0.42$) is much smaller than $|\alpha|^2=0.64$. It means that, when $\Omega=3.0\Gamma$ and $\phi=0.5\pi$, ${\sigma_{+}}$ photons are mainly converted to the $\pi$ photon and then the amplitude of the conversion photons is negative, which consequently leads to a destructive interference among the photons of different processes. When choosing other parameters as $\Omega=0.36\Gamma$ and $\phi=0.0$ in Fig.~3(c), the transmission and reflection spectra are asymmetrical.
Especially, as shown in Fig.~3(d), the total reflection probabilities of photons with different frequencies in the system will be drastically suppressed with $R_{{\sigma_{+}}}+R_{\pi}\approx0$ as the Rabi frequencies of the driving field increase to $\Omega=10\Gamma$. The insert of Fig.~3(d) points out that even when $\Omega=10\Gamma$, the transmission probabilities of the ${\sigma_{+}}$ photon are amplified with $T_{{\sigma_{+}}}>|\alpha|^2$ due to the constructive interference, which indicates that the atom is still coupled to the waveguide.
Consequently, the total transmission properties of the incident photon can be modulated to unidirectionally transport along the right direction even under non-ideal chiral photon-atom interactions.
Therefore, both the phase and the Rabi frequency of the external field can change the amplitude of the reemitted photons with frequency conversions between the two kinds of photons, and thus modify the interferences of different sources.
\begin{figure}[htbp]
  \centering
  \includegraphics[width=8cm]{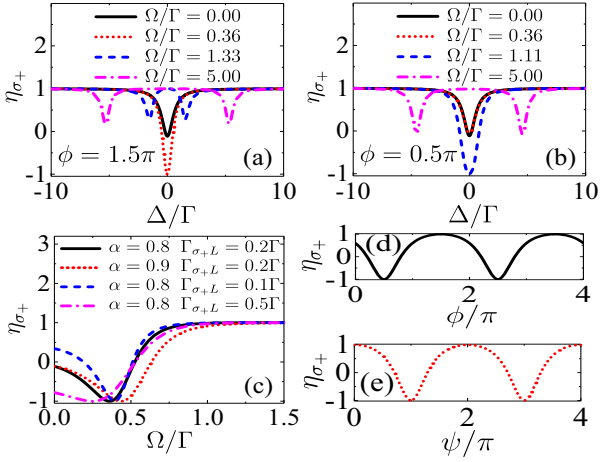}\\
  \caption { Unidirectional transmission contrasts of resonant photons under the non-ideal chiral photon-atom interactions. $\Delta=0$ in (c), (d), and (e). $\phi=1.5\pi$ in (c) and (e). $\Omega=1.11\Gamma$ in (d) and (e). Other parameters are the same as in Fig.~2. We can see that the completely unidirectional transmissions of ${\sigma_{+}}$ photon can be achieved by adjusting the Rabi frequency and the phases of the external field and of the superposition state.}\label{fig1}
\end{figure}

Now, we show how to improve the UTC in the non-ideal photon-atom interactions. As demonstrated in Fig.~4(a), without the driving field ($\Omega=0$) and in realistic experimental chiralities of $\Gamma_{{\sigma_{+}} R}/\Gamma_{{\sigma_{+}} L}=5$~\cite{MahmoodianOME43}, the UTC is relatively small. As we introduce the driving field, the ${\sigma_{+}}$ photon can be controlled to be completely reflected $\eta_{\sigma_{+}}=-1$ or transmitted $\eta_{\sigma_{+}}=1$. Although the driving field with the phase of $\phi=0.5\pi$ suppresses the maximum of the transmission probabilities, the UTC can also be adjusted from $\eta_{\sigma_{+}}=-1$ to $\eta_{\sigma_{+}}=1$, as shown in Fig.~4(b). Figure 4(c) shows that the UTC in different systems with various superposition states and chiralities can be optimized to complete transmission or reflection. Especially, as illustrated in Fig.~4(d), we can see that the completely unidirectional transport of the ${\sigma_{+}}$ photon is able to be obtained by properly choosing the phases of the driving field~\cite{SamutpraphootPRL063602}. Additionally, Fig.~4(e) indicates that the UTC can also be modulated from $\eta_{{\sigma_{+}}}=-1$ to $\eta_{{\sigma_{+}}}=1$ by choosing the phase of the superposition state $\psi$. Therefore, the $V$-type three-level atom driven by a microwave field coupled to the waveguide with non-ideal photon-atom interactions provides a controllable manner to achieve completely unidirectional single-photon quantum transport.

\begin{figure}[htbp]
  \centering
  \includegraphics[width=7.5cm]{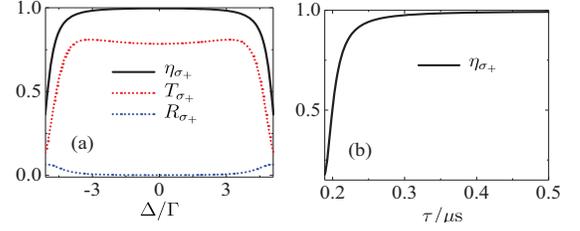}\\
  \caption {Unidirectional transmission contrast influenced by the frequency detunings between the incident single photon and the atom (a) and by the pulse length of the single photon (b). $\Omega=5\Gamma$, $\tau\approx 1/\Delta$, $\Gamma \approx \text{1MHz}$, and other parameters are the same as in Fig.~2.}\label{fig1}
\end{figure}

And then we investigate the UTC affected by the pulse length of the single photon. When the duration of the single-photon pulse is much longer than the spontaneous lifetime of the atom, the frequency of the incoming single photon can be considered to be monochromatic and the atom is treated as in the ground state~\cite{FanPRA063821}. Recent experiments about single-photon transporting can meet this precondition easily~\cite{WallraffN162,AstafievS840,HoiPRL073601,SolnerNN775}. Therefore, we can adopt the monochromatic approximation to investigate the transport properties of single photons in one-photon Fock state as denoted in Eq.~(\ref{eq:wave})~\cite{ShenPRL213001,ShenPRA023837,ZhouPRl100501,ZhouPRL103604,YanPRA045801,YuanOL5140,BradfordPRA043814}. Actually, this approximation has been commonly adopted in single photons scattered by single atoms with the analytical results~\cite{ShenPRL213001,ShenPRA023837} agreeing extremely well with the relevant single-photon transport experiments~\cite{WallraffN162,AstafievS840,HoiPRL073601,SolnerNN775}. On the other side, the advantage of having chiral coupling between photon and atom for generating quantum state transfer with high fidelities is based on the single photon resonant with the atoms~\cite{LodahlN473}. Therefore, the pulse length of the incident photon needs to be precisely prepared to satisfy the monochromatic condition. Fortunately, as shown in Fig.~5, even when the frequency detunings of the incident photon is much larger than the decaying rates, i.e., $\Delta=3\Gamma$, the UTC is still maintained to $\eta_{\sigma_{+}}=98\%$. It means that the UTC can be preserved to high values in a wide frequency range. Typically, with the atomic decaying rate to the
waveguide being $\Gamma\approx \text{1MHz}$~\cite{LodahlN473,SolnerNN775} and the pulse length being $\tau\approx1/\Delta$, as shown in Fig.~5(b), the unidirectional transmission contrast can be maintained to nearly $100\%$ when the pulse length is as long as microseconds.

\begin{figure}[htbp]
  \centering
  \includegraphics[width=8cm]{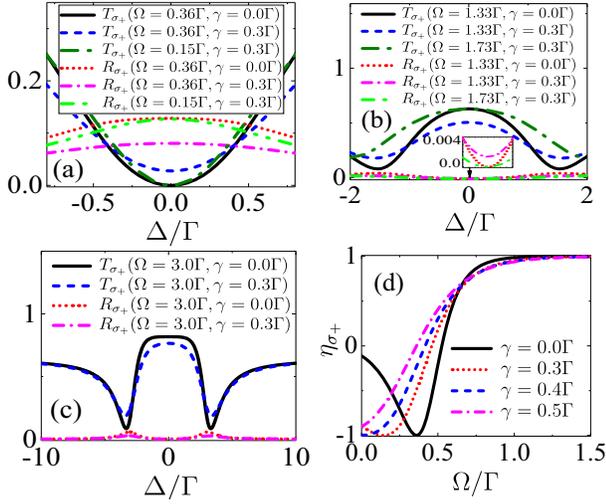}\\
  \caption {Optimizing unidirectional transmission contrast in dissipative cases, with $\Delta=0$ and $\phi=1.5\pi$. Other parameters are the same as in Fig.~2. It is obvious that the atomic dissipations into the environment can be compensated by adjusting the Rabi frequency of the driving field and the ideal UTCs can also be reached as $\eta_{\sigma_{+}}=-1$ and $\eta_{\sigma_{+}}=1$.}\label{fig1}
\end{figure}

Finally, we demonstrate the UTC against the dissipations. A recent experiment has achieved a coupling efficiency of a QD coupled to a glide-plane photonic crystal waveguide in excess of $98\%$~\cite{SolnerNN775}, and then it is safe to choose the dissipative rate of the atom as $\gamma_{\sigma_{+}}=\gamma_\pi=\gamma=0.3\Gamma$ to calculate the corresponding influence on UTC. Usually, the unavoidable intrinsic dissipative processes in the atom-waveguide system always result in the leakage of photons into non-waveguide degrees of freedom. As shown in Figs.~5(a) and 5(b), the dissipations of the reemitted photons into the environment suppress the amplitude of the reemitted photon, and thus the amplitudes of photons derived from various sources are different. These differences change the interference of the resonant photon from perfectly destructive interferences of $T_{\sigma_{+}}(\Omega=0.36\Gamma, \gamma=0.0)=0$ and $R_{\sigma_{+}}(\Omega=1.33\Gamma, \gamma=0.0)=0$ to non-perfectly destructive interferences of $T_{\sigma_{+}}(\Omega=0.36\Gamma, \gamma=0.3\Gamma)>0$ and $R_{\sigma_{+}}(\Omega=1.33\Gamma, \gamma=0.3\Gamma)>0$. By properly adjusting the Rabi frequency to change both the amplitudes of photons decaying from the ${\sigma_{+}}$ transition and of photons converted from the $\pi$ transition, the preconditions required for destructive interferences are satisfied again with $T_{\sigma_{+}}(\Omega=0.15\Gamma, \gamma=0.3\Gamma)=0$
and $R_{\sigma_{+}}(\Omega=1.73\Gamma, \gamma=0.3\Gamma)=0$. Therefore, even in the dissipative case, the complete destructive interference can be maintained as well. Especially, when the Rabi frequency of the external driving field is strong enough as $\Omega=3.0\Gamma$, as shown in Fig.~5(c), the destructive interference is insensitive to the dissipations with $R_{\sigma_{+}}(\Delta=0,\gamma=0.3\Gamma)\approx0$. Although the maximum of the transmission probabilities is suppressed as $T_{\sigma_{+}}(\Delta=0,\gamma=0.3\Gamma)<T_{\sigma_{+}}(\Delta=0,\gamma=0.0)$, the completely unidirectional transmission of $\eta_{\sigma_{+}}=1$ is still guaranteed. It can be seen directly in Fig.~5(d) that the UTC in the dissipative case can also be adjusted from $\eta_{\sigma_{+}}=-1$ to $\eta_{\sigma_{+}}=1$ by changing the values of Rabi frequency.
Particularly, the maximum dissipation required to ensure the complete destructive interference along the forward direction of $\eta_{\sigma_{+}}=-1$ is $\gamma_{\text{max}}=0.4\Gamma$, as shown  in Fig.~5(d). This maximum dissipation can be directly calculated from Eq.~(\ref{eq:tsigma}) as $\gamma=(\Gamma_{{\sigma_{+}} R}-\Gamma_{{\sigma_{+}} L})/2=0.4\Gamma$ with $\Omega=0$ and then $T_{\sigma_{+}}=0$. When the dissipation is much larger, as $\gamma=0.5\Gamma>\gamma_{\text{max}}$, the complete destructive interference is no longer preserved with $T_{\sigma_{+}}>0$ and thus $\eta>-1$.
Therefore, the completely unidirectional transmission of single photons discussed here is robust against both the non-ideal chiral photon-atom interactions and dissipations, which provides a promising way to construct cascaded quantum structures.

\section{Discussions and Conclusions}
We have studied the coherent transporting processes of single photons scattered by the single atom with non-ideal chiral photon-atom interactions. In contrast to the information carried by the incident photon inevitably reflected by the atom in non-ideal chiral structure, the results of the present work show that the ${\sigma_{+}}$ photon can transport along the waveguide with ideal unidirectional transmission by adjusting the phase and the Rabi frequency of the external driving field. Tuning the Rabi frequency, the complete destructive interference is maintained so that the UTC could be optimized to $\eta_{\sigma_{+}}=\pm1$, even when the atom possesses dissipations into non-waveguide modes. Because of the constructive interference between photons from different sources, the transmission properties of the ${\sigma_{+}}$ photons could be enhanced. In addition, the total reflection of a single photon in both modes can be simultaneously suppressed by preparing the input photon in an appropriate state. The present structure can be utilized to construct cascaded quantum networks in non-ideal chiral systems.

One of the promising candidates for experimental implementations of the above results is a QD coupled to the glide-plane photonic crystal waveguide with precisely localizing the QD near the C-point~\cite{SolnerNN775,MahmoodianOME43}. According to the whispering-gallery mode microresonator including several spatial modes with different charities, a Rb atom with fine structure driven by the magnetic field coupled to the microresonator may be adopted to realize the complete unidirection as well~\cite{RosenblumPRA033814,RosenblumNP2015,ShomroniS903}. Experimentally, the single photon in a superposition state of different frequencies can be realized with electro-optic modulators~\cite{{HePRL060501}}, wherein the mixing amplitudes of the superposition state (i.e., $\alpha$ and $\beta$) can be modulated to arbitrary values by controlling the amplitude and phase of the electric field.
Finally, when the TM mode and the TE mode are converted by an integrated mode converter, for example the tapered submicron silicon ridge optical waveguides~\cite{DaiOE13425}, and then the incident single photon in the superposition state of two photonic modes TE and TM as $\alpha|{\sigma_{+}}\rangle+\beta|{\pi}\rangle$ can be constructed. Therefore, the system described by the Hamiltonian can be realized within the current photonic technologies.

{\bf Acknowledgements.---}C.-H.Y. was supported by the National Natural Science Foundation
of China (NSFC) under Grants No.~11304210 and No.~11804240. M.L was supported by the NSFC under Grants No. 11904316. Y.-L.Z. was supported by the NSFC under Grants No. 11704370. C.-L.Z. was supported by NSFC under Grants No. 11874342 and NO. 11922411, Anhui
Initiative in Quantum Information Technologies under Grants No. AHY130000, and the Program of State Key Laboratory of Quantum Optics and Quantum Optics Devices under Grants No. KF201809.

\end{document}